\documentclass[twocolumn,showpacs,preprintnumbers,amsmath,amssymb,final]{revtex4}

\usepackage{graphicx}% Include figure files
\usepackage{dcolumn}% Align table columns on decimal point
\usepackage{bm}% bold math

\def\BEq{\begin{equation}}
\def\EEq{\end{equation}}
\def\BEqA{\begin{eqnarray}}
\def\EEqA{\end{eqnarray}}
\def\BEn{\begin{enumerate}}
\def\EEn{\end{enumerate}}
\def\BWT{\begin{widetext}}
\def\EWT{\end{widetext}}
\def\a{\alpha}

\def\b{\beta}

\def\D{\Delta}
\def\e{\epsilon}

\def\g{\gamma}
\def\G{\Gamma}

\def\w{\omega}

\def\bra{\langle}

\def\ket{\rangle}

\def\& { and }

\begin{document}

%\preprint{APS/123-QED}

\title{Resonator-zero-qubit architecture for superconducting
qubits}

\author{Andrei Galiautdinov,$^1$ Alexander N.
Korotkov,$^1$ and John M. Martinis$^2$}
 \affiliation{ $^1$Department of Electrical
Engineering, University of California, Riverside, California 92521,
USA \\
$^2$Department of Physics, University of California, Santa Barbara,
California 93106, USA}

\date{\today}% It is always \today, today,
             %  but any date may be explicitly specified

\begin{abstract}
We analyze the performance of the Resonator/zero-Qubit (RezQu)
architecture in which the qubits are
complemented with memory resonators and coupled via a resonator bus.
Separating the stored information from the rest of the processing circuit
by at least two coupling steps and the zero
qubit state results in a significant increase of the
ON/OFF ratio and  the reduction of the idling error. Assuming no decoherence,
we calculate such idling error, as well as the errors for the MOVE operation
and tunneling measurement, and show that the RezQu architecture can provide
high fidelity performance required for medium-scale quantum information
processing.

\end{abstract}


%\keywords{Suggested keywords}%Use showkeys class option if keyword
                              %display desired
\maketitle

%\tableofcontents

\section{Introduction}

Superconducting circuits with Josephson junctions are steadily
gaining attention as promising candidates for the realization of a
quantum computer \cite{CLARKE2008}. Over the last several years,
significant progress has been made in preparing, controlling, and
measuring the macroscopic quantum states of such circuits
\cite{Yamamoto03,Plantenberg07,Clarke-06,Steffen06,DiCarlo09,DiCarlo10,
Mariantoni-Science,Simmonds-10,Oliver-11,Esteve-09,Siddiqi-11}.
 However, the two major roadblocks -- scalability and decoherence
-- still remain, impeding the development of a workable prototype.
The Resonator/zero-Qubit (RezQu) protocol presented here aims to
address these limitations at a low-level (hardware cell)
architecture \cite{DiVincenzo}.

\begin{figure}
\includegraphics[angle=0,width=1.00\linewidth]{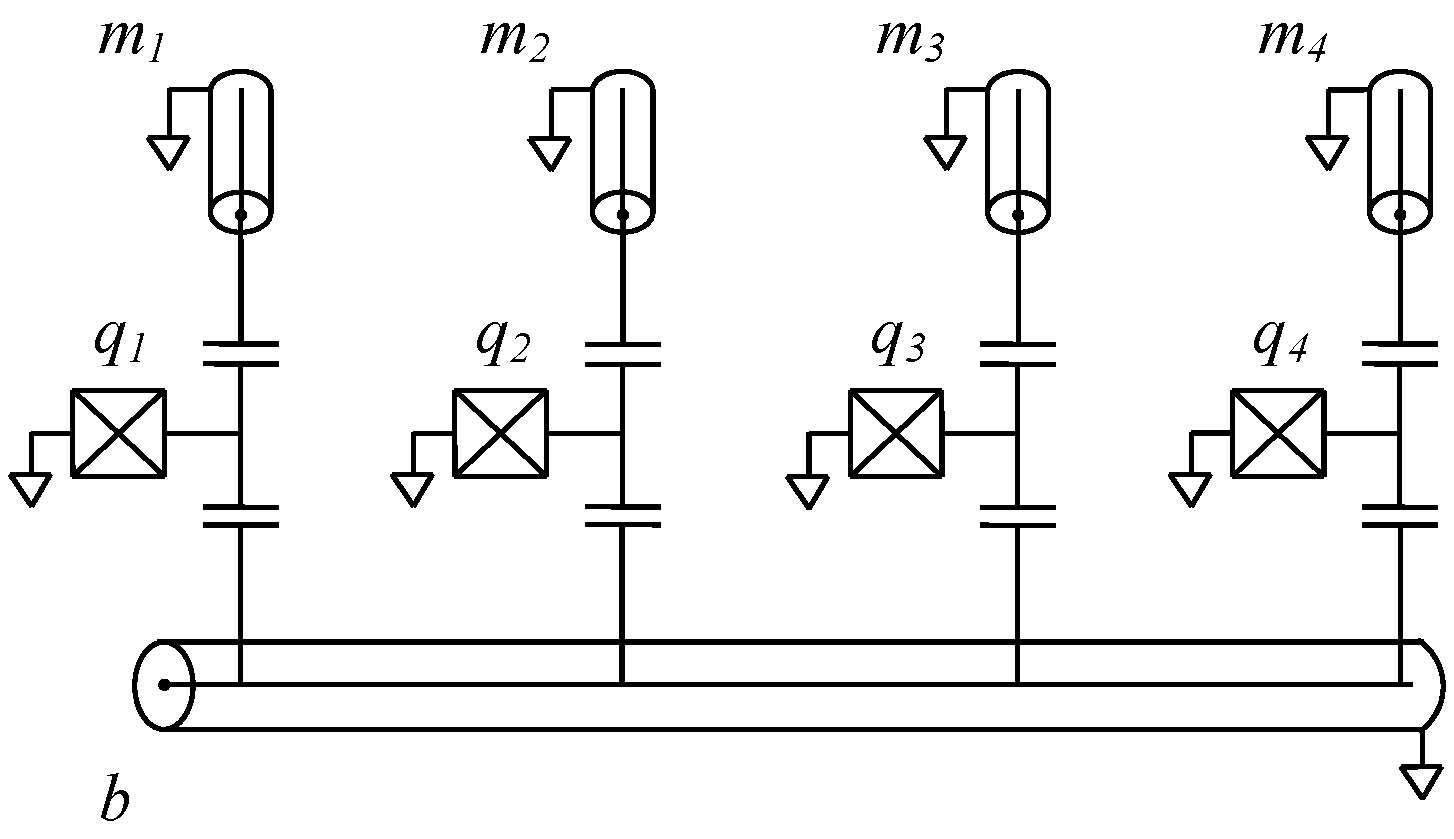}
\caption{ \label{fig:1} Schematic diagram of the RezQu architecture:
$m$ -- memory resonators, $q$ -- qubits, $b$ -- bus. We assume frequencies
$\sim$7 GHz for the memories, $\sim$6 GHz for the bus, and qubit frequencies
are varied between these values.}
\end{figure}

A RezQu device consists of a set of superconducting qubits (e.g., phase
qubits \cite{Martinis-QIP}), each of which is capacitively coupled to its
own memory resonator and also capacitively coupled to a common
resonator bus, as shown in Fig.\ \ref{fig:1}
\cite{WANG2011,Mariantoni-NatPhys,Mariantoni-Science,WILHELM_NOON_2010}.
The bus is used for coupling operations between qubits, while the memory
resonators are used for information storage when the logic qubits
are idling. With coupling capacitors being fixed and relatively
small, qubit coupling is adjusted by varying the qubit frequency,
which is brought in and out of resonance with the two resonators.  For a
one-qubit operation, quantum information is moved from the memory to
the qubit, where a microwave pulse is applied. A natural two-qubit
operation is the controlled-$Z$ gate, for which one qubit state is
moved to the bus, while the other qubit frequency is tuned close to
resonance with the bus for a precise duration
\cite{Strauch03,Haack10,DiCarlo10,Yamamoto10}.

Most importantly, the information stored in resonators is separated
from the rest of the processing circuit by the known qubit state
$|0\ket$ and at least two coupling steps, thus reducing crosstalk
error during idling.
  Also, the problem of spectral crowding is essentially eliminated
because the two-step resonance between empty qubits is not harmful,
while the four-step coupling between memory resonators is
negligible. Therefore the resonator frequencies, which are set by
fabrication, can be close to each other, decreasing sensitivity to
phase errors in the clock.
  Thus the RezQu architecture essentially solves the inherent ON/OFF ratio
problem of the {\it fixed} capacitive coupling without using a more
complicated scheme of a tunable coupling
\cite{Clarke-06,Nakamura-07-tunable,BIALCZAK2010}.
 As an additional benefit, information storage in
resonators increases coherence time compared to storage in the
qubits.
 We note that the idea of using resonators to couple qubits has been
suggested by many authors
\cite{Zhu03,Blais03,Zhou04,Blais04,Cleland04,Blais07,
Koch07,MAJER2007,Girvin08,NORI11}. The use of
resonators as quantum memories has also been previously proposed
\cite{Pritchett05,Silanpaa07,Girvin08,Johnson10}. However, putting
the two ideas together in a single architecture results in the new
qualitative advantages, which have not been discussed.

%%%%%%%%% BEGIN FIG. 1
%%%%%%%%% END FIG. 1

In this paper we briefly consider the relation between the logical
and the physical qubit states and then analyze several basic
operations in the RezQu architecture. In particular, for a truncated
three-component memory-qubit-bus RezQu device we focus on the idling
error, information transfer (MOVE) between the qubit and its memory,
and the tunneling measurement. The analysis of the controlled-$Z$
gate will be presented elsewhere \cite{JoydipGhosh11}. For
simplicity decoherence is neglected.

%\section{Error analysis of the RezQu architecture}

\section{Logical vs physical qubits}
%\label{sec:logical_vs_actual}

%\subsection*{Logical vs physical qubits}

 We begin by recalling an important difference between logical and actual,
physical qubits. The difference stems from the fact that in the
language of quantum circuit diagrams the idling qubits are always
presumed to be stationary \cite{NIELSEN&CHUANG}, while
superconducting qubits evolve with a 6-7 GHz frequency even in
idling, which leads to accumulation of relative phases. Also, the
always-on coupling leads to fast small-amplitude oscillations of the
``bare''-state populations during the off-resonant idling. A natural
way to avoid the latter problem is to {\it define} logical
multi-qubit states to be the (dressed) {\it eigenstates} of the
whole system (see\, e.g.,
\cite{Blais04,Blais07,Koch07,Boissonneault09,DiCarlo09,Pinto10}  for
discussion of the dressed states).
    Then the only evolution in idling is the phase accumulation for
each logic state. However, there are $2^N$ logic states for $N$
qubits, and using $2^N$ ``local clocks'' (rotating frames) would
require an exponential overhead to calibrate the phases. The
present-day experiments with two or three qubits often use this
unscalable way, but it will not work for $N\agt 10$. A scalable way
is to choose only $N$ rotating frames, which correspond to one-qubit
logical states, and treat the frequencies of multi-qubit states only
approximately, as sums of the corresponding single-qubit
frequencies. The use of such non-exact rotating frames for
multi-qubit states leads to what we call an idling error, which is
analyzed in the next section.

 Notice that it is sufficient to establish a correspondence
between logical and physical states only at some time moments
between the gates. Moreover, this correspondence may
be different at different moments. At those moments, the bus is
empty, the $2N+1$ system components ($N$ qubits, $N$ memories, and the bus)
are well-detuned from their neighbors, and for each logical qubit it is
unambiguously known whether the corresponding quantum information is
located in the memory or in the qubit.
Thus the eigenstates corresponding to the logical states are
well-defined and the physical-to-logical correspondence is
naturally established by projecting onto the $2^N$ computational eigenstates,
while occupations of other eigenstates should be regarded as an error.
In the simplest modular construction of an algorithm we should not attempt
to correct the error of a given gate by the subsequent gates. Then for the
overall error we only need to characterize the errors of individual gates,
as well as the idling error.

    One may think that defining logical states via the eigenstates of the
whole system may present a technical problem in an algorithm design.
However, this is not really a problem for the following reasons.
First, we need the conversion into the basis of eigenstates only at
the start and end of a quantum gate, while the design of a gate is
modular and can be done using any convenient basis. Second, in
practice, we can truncate the system to calculate the eigenstates
approximately, making sure that the error due to truncation is
sufficiently small. Similar truncation with a limited error is
needed in practical gate design.

    As mentioned above, the physical-to-logical correspondence rule
can be different at every point between the gates. For the
correspondence based on eigenstates we are free to choose $N$
single-excitation phases arbitrarily. In spite of this freedom, for
definiteness, it makes sense to relate all the single-excitation
phases to a particular fixed time moment in an algorithm. Then a
shift of the gate start time leads to easily calculable phase
shifts, which accumulate with the frequencies equal to the change of
single-excitation frequencies before and after the gate. Such shift
is useful for the adjustment of relative single-excitation phases
\cite{Hofheinz}. Another way of the single-excitation phase
adjustment is by using ``qubit frequency excursions''
\cite{Mariantoni-Science}. The ease of these adjustments
significantly simplifies design of quantum gates, because we
essentially should not wary about the single-excitation phases.

    Notice that initial generation of high-fidelity single-excitation
eigenstates is much easier experimentally than generation of the
bare states. This is because a typical duration of the qubit
excitation pulses is significantly longer than inverse detuning
between the qubit and resonators. We have checked this advantage of
using eigenstates versus bare states numerically in a simple model
with typical parameters \cite{Mariantoni-Science} of a RezQu device
(the error decrease is about two orders of magnitude). Similarly,
the standard one-qubit operations essentially operate with the
eigenstates rather than with the bare states.

%[Note that various applications of dressed states to circuit QED have
%been presented in the following Refs.: \cite{Blais04}, where a second order
%and exact corrections were given for a qubit-cavity coupling,
%\cite{Boissonneault09}, where a fourth
%order correction was considered,
%\cite{Blais07}, where a second order correction for two qubits coupled to a
%resonator was discussed,
%\cite{Koch07} and \cite{DiCarlo09}, in which, respectively,  a second and fourth order
%corrections taking into account multi-levels of the atoms were analyzed.]

\section{Idling error}

%\subsection*{Idling error}

    Before discussing the idling error in the RezQu architecture, let us consider
a simpler case of two directly coupled qubits. Then in idling the
wavefunction evolves as $|\psi(t)\ket = \a_{00} e^{-i\e_{00}t}
|\overline{00}\ket + \a_{01} e^{-i\e_{01}t}
|\overline{01}\ket+\a_{10} e^{-i\e_{10}t} |\overline{10}\ket +
\a_{11} e^{-i\e_{11}t} |\overline{11}\ket$, where we denote the
logical (eigen)states with an overline, their corresponding
(eigen)energies with $\e_{ij}$, and amplitudes at $t=0$ with
$\a_{ij}$. However, for the desired evolution $|\psi_{\rm
desired}(t)\ket$ the last term should be replaced with
$\a_{11}e^{-i(\e_{01}+\e_{10} -\e_{00})t} |\overline{11}\ket$; then
only two rotating frames (clocks) with frequencies $\e_{01}-\e_{00}$
and $\e_{10}-\e_{00}$ are needed. We see that the phase difference
accumulates with the frequency $\Omega_{ZZ} =
(\varepsilon_{11}-\varepsilon_{01})-
(\varepsilon_{10}-\varepsilon_{00})$, and therefore the idling error
due to qubit coupling accumulates over a time $t$ as
    \begin{align}
{\rm Err}
& = 1-|\bra \psi_{\rm desired}(t)| \psi(t)\ket|^2
    \nonumber \\
& \simeq |\a_{11}|^2(1-|\a_{11}|^2)(\Omega_{ZZ}\,t)^2 \alt (\Omega_{ZZ}\,t)^2 ,
\label{idleerror} \end{align}
where we assumed $\Omega_{ZZ}\,t \ll 1$. (The frequency $\Omega_{ZZ}$
is defined in the same way as in Ref.\ \cite{Pinto10} for
a two-qubit $\sigma_{Z}^{(1)}\sigma_Z^{(2)}$ interaction.)
 The error is state-dependent, but in
this paper we will always consider error estimates for the worst-case scenario.

In the RezQu architecture, the main contribution to the idling error comes
from interaction between a memory resonator, in which quantum information is stored,
and the bus, which is constantly
used for quantum gates between other qubits. By analogy with the above case, for
the truncated memory-qubit-bus ($mqb$) system the idling error can be estimated as
    \BEq
\label{eq:idlingErrorFrequency}
{\rm Err} \simeq (\Omega_{ZZ}\,t)^2, \,\,\,
\Omega_{ZZ} =  (\e_{101} - \e_{001}) - (\e_{100} - \e_{000}),
    \EEq
where the eigenenergies correspond to the logical eigenstates
$|\overline{101}\ket$, $|\overline{001}\ket$, $|\overline{100}\ket$,
$|\overline{000}\ket$, and in our $|\overline{mqb}\ket$ notation the
sequence of symbols represents the states of the memory resonator,
the qubit, and the bus. Notice that the qubit here is always in
state $|0\ket$, and $\Omega_{ZZ}$ is essentially the difference
between the effective frequencies of the memory resonator in the
presence and absence of the bus excitation.

    To find $\Omega_{ZZ}$ we use the rotating wave approximation (RWA); then
the dynamics of the $mqb$ system is described by the Hamiltonian
(we use $\hbar=1$)
\begin{align}
\label{RWAhamiltonian1}
H(t) &=
\begin{bmatrix}
0 & 0 & 0\cr
0 & \w_q(t) &0 \cr
0&0& 2\w_q(t) - \eta
\end{bmatrix}
+ \w_m a^{\dagger}_m a_m
+ \w_b a^{\dagger}_b a_b
\nonumber \\
& + g_m
\left(a^{\dagger}_m \sigma_{q}^{-} + a_m\sigma_{q}^{+} \right)
+ g_b
\left(\sigma_{q}^{-}a^{\dagger}_b + \sigma_{q}^{+}a_b \right),
\nonumber \\
& + g_{d}
\left(a^{\dagger}_m a_b + a_m a^{\dagger}_b\right),
\end{align}
where the qubit frequency $\w_q$ may vary in time, while
the qubit anharmonicity $\eta$ is assumed to be constant,
\BEq
\sigma_{q}^{-} =
\begin{bmatrix}
0 & 1 &0\cr
0 & 0 &\sqrt{2} \cr
0 & 0 &0
\end{bmatrix},
\quad \sigma_{q}^{+} = \left(\sigma_{q}^{-}\right)^{\dagger}
  \EEq
are the qubit lowering and raising operators, $\w_m$, $\w_b$ are the
memory and the bus frequencies (which are presumed to be fixed),
$a^{\dagger}_m$, $a_m$, $a^{\dagger}_b$, $a_b$ are the
creation/annihilation operators for the memory and the bus photons,
and $g_m$, $g_b$ are the memory-qubit and qubit-bus coupling
constants. The last term in Eq.\ (\ref{RWAhamiltonian1}) describes
the direct (electrostatic) memory-bus coupling; replacing a qubit in
Fig.\ 1 with a lumped tank circuit it is found
\cite{PRYADKO&KOROTKOV_iSWAP} to be
    \BEq
    g_d = 2g_mg_b/\w_q.
    \label{g_d}\EEq
It is typically smaller than the effective memory-bus coupling via the virtual
excitation of the qubit because the detunings $|\omega_m-\omega_q|$ and
$|\omega_b-\omega_q|$ between the elements are much smaller than their
frequencies; because of that, we often neglect $g_d$. From the physical model
it is easy to show that $g_m$ and $g_b$ are proportional to $\omega_q^{1/2}$
and therefore change when the qubit frequency is varied; however for simplicity
we will assume constant $g_m$ and $g_b$.

Neglecting $g_d$, in fourth order we find (see Appendix),
    \begin{eqnarray}
\label{eq:OmegaZZ_4th_order}
&& \Omega_{ZZ} =  \frac{-2g_m^2 g_b^{2}\eta}
{\Delta_m^2\Delta_b^2} \,
\frac{\omega_m+\omega_b-2\omega_q}{\omega_m+\omega_b-(2\omega_q-\eta)},
    \\
&&    \Delta_m=\omega_m-\omega_q, \,\,\, \Delta_b=\omega_q-\omega_b,
    \end{eqnarray}
which is very close to the exact value found by direct
diagonalization of the Hamiltonian (Fig.\ \ref{fig:2}), and the
effect of $g_d$ is of a higher order and therefore very small (see
Appendix and Fig.\ \ref{fig:2}). Notice that $\Omega_{ZZ}\propto
\eta$ because in a linear system $\Omega_{ZZ}=0$, and nonlinearity
comes from the qubit. Equation (\ref{eq:OmegaZZ_4th_order}) shows
that an optimal choice of the qubit ``parked'' frequency is $\w_q =
(\w_m+\w_b)/2$, midway between the memory and the bus frequencies;
then the idling error in this order goes to zero (this happens
because the contribution of $|020\rangle$ in
$|\overline{101}\rangle$ becomes zero -- see Appendix).
    Notice that in the RezQu architecture the frequencies of the memory
resonators are assumed to be relatively close to each other (forming a ``memory
band'' of frequencies). Then the optimal ``parked'' frequencies of the qubits
are also close to each other. This is not a problem when all qubits
are in state $|0\ket$; however, when a qubit is excited this may lead
to a significant resonant coupling with another qubit via the bus. To avoid
this ``spectral crowding'' effect,
it is useful to reserve two additional frequencies, situated sufficiently far
from the ``parked'' frequencies, at which a pair of qubits may undergo
local rotations (simultaneous rotations of two qubits are often useful
before and after two-qubit gates).

%%%%%%%%%%%% FIGURE 2 %%%%%%%%%%%
\begin{figure}
\includegraphics[angle=0,width=1.00\linewidth]{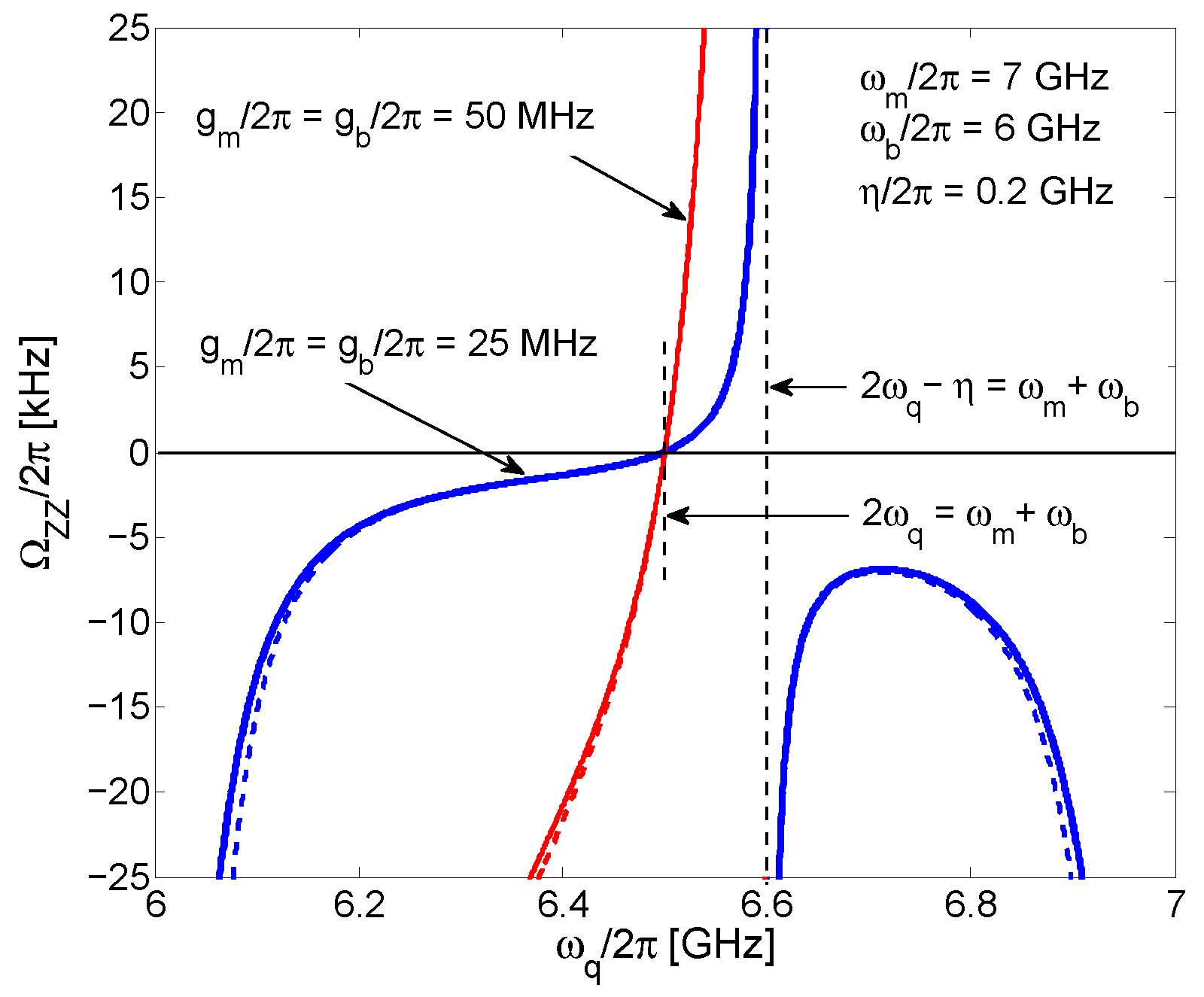}
\caption{ \label{fig:2} (Color online) The frequency $\Omega_{ZZ}$
for a truncated memory-qubit-bus system [the idling error is
$(\Omega_{ZZ}t)^2$] for two values of the coupling: $g_m/2\pi =
g_b/2\pi = 25$ MHz (blue lines) and $g_m/2\pi = g_b/2\pi = 50$ MHz
(red lines).
   The solid lines show the results of exact diagonalization of the RWA
Hamiltonain (\ref{RWAhamiltonian1}) with and without $g_d$. The
effect of $g_d$ is not visible (smaller than the line thickness).
The blue and red dashed lines show the analytical result
(\ref{eq:OmegaZZ_4th_order}). }
\end{figure}
%%%%%%%%%%%% END FIG. 2 %%%%%%%%%%

    The idling error (\ref{eq:idlingErrorFrequency}) scales quadratically
with time. This is because we use a definition for which not the error itself
but its square root
corresponds to a metric, and therefore for a composition of quantum gates in the
worst-case scenario we should sum square roots of the errors
\cite{NIELSEN&CHUANG}. For the same reason, the worst-case idling error
scales quadratically, ${\rm Err}\propto N^2$, with the number $N$ of qubits in
a RezQu device. In principle, an average idling error may scale linearly
with $N$ and time (for that we would need to define the memory ``clock''
frequency using an average occupation of the bus); however, here we use
only the worst-case analysis.

    It is convenient to replace the time-dependence in the idling error
estimate $(N\Omega_{ZZ}\, t)^2$ by the dependence on the number of
operations $N_{\rm op}$ in an algorithm. (The corresponding
quadratic dependence on $N_{\rm op}$ can also be interpreted as the
worst-case-scenario error for a composition of quantum operations.)
Assuming that each operation crudely takes time $t_{\rm op}\simeq
g_m^{-1}+g_b^{-1}$ (this estimate comes from MOVE operations
discussed later and also from controlled-$Z$ gate) and neglecting
the second factor in Eq.\ (\ref{eq:OmegaZZ_4th_order}) (i.e.\
assuming non-optimal ``parked'' qubit frequencies), we obtain the
following estimate for the worst-case idling error:
    \BEq
\label{eq:IE-RezQu}
{\rm Err} \simeq
 \frac{g_m^3g_b^3 \eta^2 N^2 N^2_{\rm op}}{\D_m^4\D_b^4}
\, \frac{{\rm max}(g_m, g_b)}{{\rm min}(g_m,g_b)},
    \EEq
where $\D_{m}$ and $\D_{b}$ are typical detunings at idling.
Using for an estimate
$g_m/2\pi=g_b/2\pi= 25$ MHz, $\eta/2\pi=200$ MHz, and $\Delta_m/2\pi
=\Delta_b/2\pi= 500$ MHz, we obtain ${\rm Err}\simeq 10^{-8} N^2 N_{\rm op}^2$.

    To demonstrate the advantage of the RezQu architecture, Eq.\
(\ref{eq:IE-RezQu}) may be compared with the corresponding result
for the conventional bus-based architecture (without additional
memories). Then the idling error is due to $ZZ$-interaction between
the qubit and the bus: the frequency of an idling qubit is affected
by the bus occupation due to logic operations between other qubits.
In this case $\Omega_{ZZ,{\rm
conv}}=-2g_b^2\eta/[\Delta_b(\Delta_b-\eta)]$, that gives
    \BEq \label{eq:IE-standard}
    {\rm Err}_{\rm conv} \simeq
g_b^2 \eta^2 N^2 N^2_{\rm op}/\D_b^4.
    \EEq
    Assuming a typical ratio
$g/\Delta \lesssim 0.1$ between the coupling and detuning, we have a
reduction in the idling error in the RezQu architecture by at least
$10^4$ even before considering that $\Omega_{ZZ}$ can be zeroed in
this order. Using Eq.\ (\ref{eq:IE-standard}) we see that a
conventional architecture allows only a very modest number of qubits
and operations before the idling error becomes significant. In
principle, the problem can be solved by a constantly running
dynamical decoupling (which would be quite nontrivial in a
multi-qubit device). The RezQu idea eliminates the need for such
dynamical decoupling.

    All our estimates so far were for the idling error due to the
memory-bus interaction. Now let us discuss errors due to the
four-step memory-memory interaction in the RezQu architecture. The
$XX$-interaction between the memory and another ($k$th) memory can
be calculated \cite{Pinto10} as
$\Omega_{XX}=2g_{m}g_{m,k}g_{b}g_{b,k}/[\Delta_{m}\Delta_{m,k}
    (\omega_m-\omega_b)]$, where additional subscript $k$ indicates
parameters for the $k$th section of the device. The $XX$-interaction
does not produce a phase error accumulating in idling, but leads to
the error ${\rm Err}\simeq (\Omega_{XX}/\delta_m)^2$ every time the
information is retrieved from memory, where $\delta_m$ is the
typical spacing between memory frequencies. Assuming similar
sections of the RezQu device with $\Delta_b\simeq\Delta_m\equiv
\Delta$ and $\delta_m\simeq \Delta/N$, we obtain an error estimate
${\rm Err} \simeq (Ng_m^2 g_b^2/\Delta^4)^2$ per operation. Since
the worst-case scaling with the number of operations $N_{\rm op}$ is
always $\propto N_{\rm op}^2$, we obtain the worst-case estimate
    \BEq
    {\rm Err} \simeq N^2 N_{\rm op}^2g_m^4 g_b^4 /\Delta^8.
    \label{Err-m-m1}\EEq
This error is smaller than the idling error (\ref{eq:IE-RezQu}) if
$g_{m,b}<\eta$. For $g_m= g_b= \Delta /20$ we find a very small error
${\rm Err}\simeq 10^{-10} N^2N_{\rm op}^2$, which means that there is
essentially no spectral crowding problem for memories.
 Notice that for a conventional bus-based architecture the
error estimate (\ref{Err-m-m1}) is replaced by ${\rm Err_{conv}}\simeq
N^2 N_{\rm op}^2 g_b^4 /\Delta^4$ and presents a difficult scaling
problem due to the spectral crowding.

    Besides the $XX$-interaction between the two memories, there is also
the $ZZ$-interaction. Using the same approximate derivation as in
Appendix [see Eq.\ (\ref{omega_zz-eta})] and assuming
$\Delta_b\simeq\Delta_m\equiv \Delta$, we find an estimate
$\Omega_{ZZ}\approx -\eta g_m^4 g_b^4/\Delta^8$. Then using ${\rm
Err}\simeq (\Omega_{ZZ} t_{\rm op})^2 N_{\rm op}^2 N^4$ (the scaling
$N^4$ is because each pair brings a contribution), with $t_{\rm
op}\simeq g_m^{-1}+ g_{b}^{-1}$, we obtain the worst-case estimate
    \BEq
{\rm Err} \simeq
 \frac{g_m^7g_b^7 \eta^2 N^4 N^2_{\rm op}}{\D^{16}}
\, \frac{{\rm max}(g_m, g_b)}{{\rm min}(g_m,g_b)}.
    \label{Err-m-2}\EEq
This error is smaller than the memory-bus idling error (\ref{eq:IE-RezQu})
if $N<\Delta^4/g_m^2g_b^2$, which is always the case in practice.

\section{MOVE operation}
%\subsection*{MOVE operation}

Any logic gate in the RezQu architecture requires moving quantum
information from one system element (memory, qubit, bus) to another.
Therefore the MOVE operation is the most frequent one.
 It is important to mention that the one-way MOVE
operation \cite{PRYADKO&KOROTKOV_iSWAP} is easier to design than the
SWAP (or $i$SWAP) operation because we are not interested in the
fidelity of the reverse transfer and can also assume zero occupation
of the neighboring element. For example, for a perfect
qubit$\rightarrow$memory MOVE ($i$MOVE) operation in the truncated
$mqb$ system we search for a unitary, which transforms
$|\overline{010}\ket \rightarrow -i |\overline{100}\ket$ (notice
that in RWA $|\overline{000}\ket \rightarrow |\overline{000}\ket$
always), but we are not interested in what happens to the initial
states $|\overline{001}\ket$  and $|\overline{100}\ket$. Moreover,
we can allow for an arbitrary phase, $|\overline{010}\ket
\rightarrow -i e^{-i\varphi} |\overline{100}\ket$, because this
phase can be compensated either by shifting the operation start time
within one period of the initial memory-qubit detuning
\cite{Hofheinz} or by ``qubit frequency excursion'' with proper
integral \cite{Mariantoni-Science}. Therefore, we need to satisfy
only two (complex) equations to design the unitary $U_{\rm MOVE}$
for this MOVE,
    \BEq
  \langle\overline{010}|U_{\rm MOVE}|\overline{010}\rangle =0,  \,\,\,
    \langle\overline{001}|U_{\rm MOVE}|\overline{010}\rangle =0.
    \label{Umove}\EEq

Experimentally the qubit$\rightarrow$memory MOVE is done
\cite{Hofheinz,WANG2011,Mariantoni-NatPhys,Mariantoni-Science} by
tuning the qubit in resonance (with some overshoot) with the memory
resonator approximately for a duration $\pi/2g_m$. Equation
(\ref{Umove}) means that any reasonable shape of $\omega_q(t)$
tune/detune pulse with four adjustable parameters can be used for a
perfect MOVE operation in the truncated $mqb$ system. Actually, as
will be discussed later, the use of only two adjustable parameters
is sufficient to obtain an exponentially small error in the
quasi-adiabatic regime. Such two-parameter construction is most
convenient for practical purposes, but formally it is imperfect
(non-zero error). So we will first discuss the perfect (zero error)
four-parameter construction.

    We have designed the qubit$\rightarrow$memory MOVE pulses $\omega_q(t)$
for the truncated $mqb$ device both
analytically (in first order) and numerically.  The initial and final
frequencies of the qubit are allowed to be different. In the analytical
design we do calculations in the bare basis,
$|\psi(t)\ket = \a(t)|100\ket + \b(t)|010\ket + \g(t)|001\ket$,
but define the co-moving frame as
  \BEq
\tilde\a = \a\, e^{i\w_m t} , \,\,\, \tilde\b = \b\, e^{i\int_0^t \w_q(t') dt'},
\,\,\,   \tilde\g = \g\, e^{i\w_b t}.
  \EEq
In this representation the only interesting initial state $|\overline{010}\ket$
of the qubit$\rightarrow$memory MOVE is (in first order)
    \BEq
\tilde\a(0) = -g_m/\D_{m}(0), \,\, \tilde\b(0) = 1, \,\,
\tilde\g(0) = g_b/\D_{b}(0),
    \EEq
where $\D_{m} = \w_m-\w_{q}$ and $\D_{b} = \w_{q}-\w_b$ are the detunings.
The desired (target) final state at time $t_{\rm f}$ is
$-i e^{-i\varphi} |\overline{100}\ket$, i.e.
\begin{align}
\label{eq:finStateMoveCoMoving}
\tilde\a(t_{\rm f}) = & -ie^{-i\varphi}, \,\,\, \tilde\g(t_{\rm f}) = 0,
\nonumber \\
\tilde\b(t_{\rm f}) = & -i e^{-i\varphi} g_m \D_{m}^{-1}(t_{\rm f}) \,
e^{-i\int_{0}^{t_{\rm f}} \D_m(t') dt'},
\end{align}
Notice that even though the phase $\varphi$ is arbitrary, the relative phase
between $\tilde\a(t_{\rm f})$ and $\tilde\b(t_{\rm f})$ is fixed by the
absence of the relative phase between $\a(t_{\rm f})$ and $\b(t_{\rm f})$.
  We see that the MOVE operation should eliminate the initial ``tail''
$\tilde\g(0)$ on the bus
 (this needs two real parameters in the pulse design)
and transfer most of the excitation to the memory with correct
magnitude and relative phase of $\tilde\b(t_{\rm f})$ (two more real
parameters).

  Similarly to the experimental pulse design
\cite{Hofheinz,WANG2011,Mariantoni-NatPhys,Mariantoni-Science},  we
assume that the shape of $\omega_q(t)$ pulse consists of a front
ramp, rear ramp, and a flat part in between them (Fig.\ \ref{fig:4}
illustrates a piecewise-linear construction of the pulse). As will
be shown below, using two parameters for the front ramp shape we can
ensure elimination of the ``tail'' $\tilde\gamma(t_{\rm f})$; we can
choose a rather arbitrary rear ramp, and using two parameters for
the flat part (its frequency overshoot and duration) we can provide
proper $\tilde\beta (t_{\rm f})$.

%%%%%%%%%%%% FIGURE 4 %%%%%%%%%%%
\begin{figure}
\includegraphics[angle=0,width=1.00\linewidth]{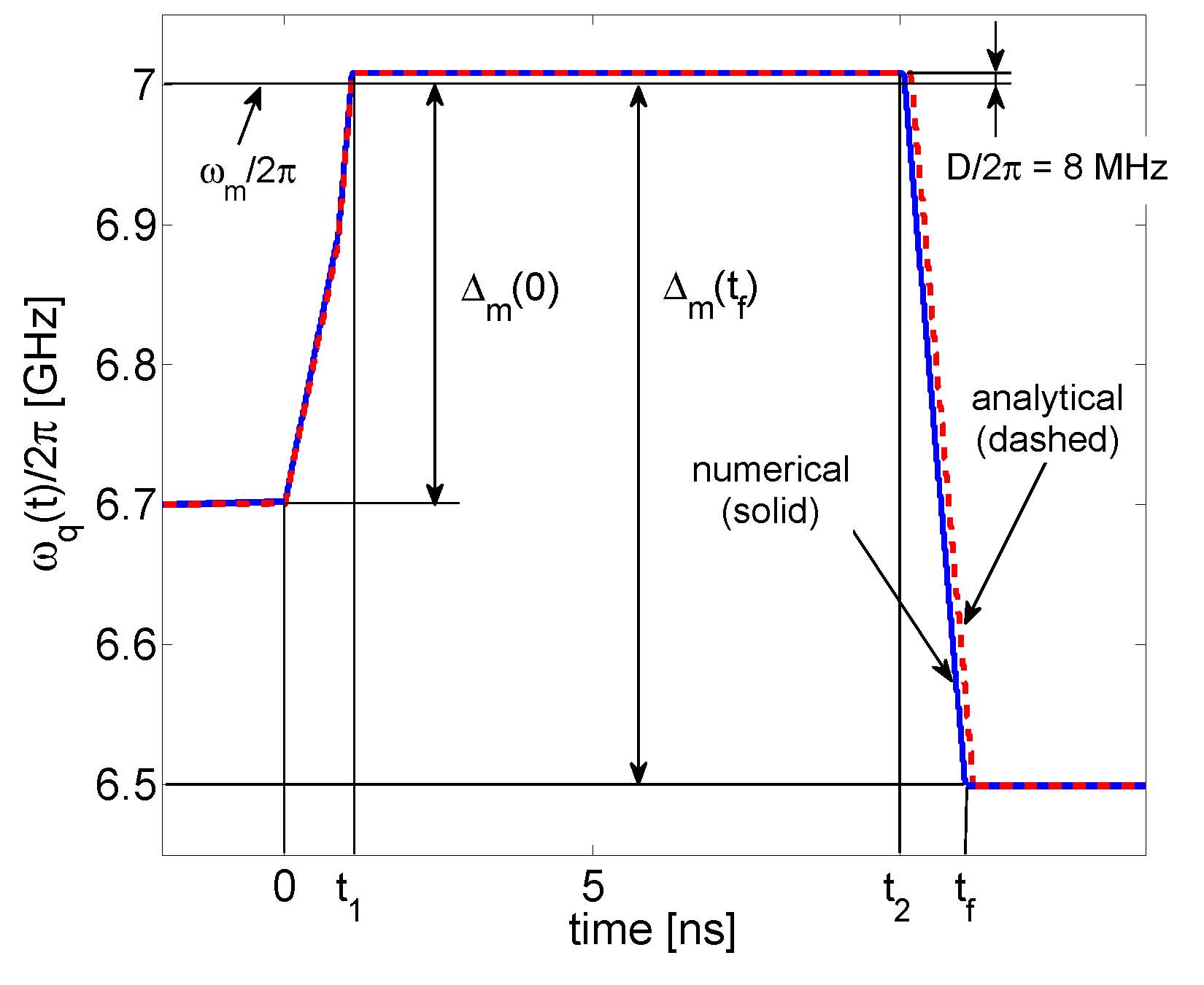}
\caption{ \label{fig:4} (Color online) Illustration of a
piecewise-linear tune/detune pulse shape (qubit frequency as a
function of time) for the MOVE operation qubit$\rightarrow$memory in
a three-component $mqb$ system. The front ramp consists of two
straight segments. The solid blue line shows the result of a
four-parameter numerical optimization, in which the slope of the
first straight segment, the qubit frequency at the end of the first
straight segment, the duration $t_2-t_1$ of the flat part and its
overshoot $D$ have been optimized. This gives ${\rm Err} = 0$ up to
machine accuracy. The red dashed line shows analytical design based
on Eqs.\ (\ref{eq:shaping2}) and (\ref{eq:overshoot}); in this case,
${\rm Err} = 5\times 10^{-4}$. System parameters: $\w_m/2\pi = 7$
GHz, $\w_b/2\pi = 6$ GHz, $\w_q(0)/2\pi = 6.7$ GHz, $\w_q(t_{\rm
f})/2\pi = 6.5$ GHz, $g_m/2\pi = g_b/2\pi = 25$ MHz, the slopes of
the second front segment and of the final ramp have been fixed at
500 MHz/ns. }
\end{figure}
%%%%%%%%%%%% END FIG. 4 %%%%%%%%%%

    Let us start with the ``tail'' $\tilde\gamma$. As follows from the
Schr\"{o}dinger equation with the Hamiltonian
(\ref{RWAhamiltonian1}), \BEq \tilde\g(t_{\rm f}) = \tilde\g(0) -
ig_b\int_{0}^{t_{\rm f}}dt \, \tilde\b(t)\, e^{-i\int_0^t
\D_b(t')dt'}. \label{tilde-gamma-1}\EEq Let us denote the end of the
front ramp by $t_1$ and the start of the rear ramp by $t_2$ (see
Fig.\ 4). For $0<t<t_1$ in Eq.\ (\ref{tilde-gamma-1}) we can replace
$\tilde\beta(t)$ with $\tilde\beta (0)=1$ because the qubit
occupation cannot change much during a short ramp. For $t_1<t<t_2$
we can use integration by parts using $\Delta_b(t)\approx
\omega_m-\omega_b$, with $\tilde\beta(t)$ changing approximately
from 1 to 0. Finally, there is a negligible (second-order in $g_b$)
contribution to the integral for $t_2<t<t_{\rm f}$ because
$\tilde\beta (t)$ is already small (first order in $g_b$). Thus for
the desired pulse shape, in first order we obtain
\begin{eqnarray}
\label{eq:shaping2}
&& 0= \frac{\tilde\g(t_{\rm f})}{g_b}
= \frac{1}{\Delta_b(0)}
- i \int_{0}^{t_{1}} e^{-i {\cal A}_0^t } dt
- \frac{e^{-i {\cal A}_0^{t_1}}}{\w_m-\w_b},\qquad
    \\
&& {\cal A}_{t'}^{t''} \equiv \int_{t'}^{t''}\D_{b}(t)\, dt.
    \label{cal-A}\end{eqnarray}
As we see, required elimination of the ``tail'' $\tilde\gamma$ on
the bus gives two equations (real and imaginary parts) for the front
ramp shape. This can be done by using practically any shape with two
adjustable parameters. Notice that condition (\ref{eq:shaping2})
essentially means that in order to have correct (zero) ``tail'' on
the bus at final time $t_{\rm f}$, this tail at time $t_1$ should be
the same as the tail for the co-moving eigenstate of the qubit-bus
system.

Now let us design the flat part of the pulse, which should give us the
proper ratio $\tilde\b(t_{\rm f})/\tilde\a(t_{t_{\rm f}})$ from Eq.\
(\ref{eq:finStateMoveCoMoving}). After designing the front ramp we know
$\tilde\alpha$ and $\tilde\beta$ at the start of the flat part $t_1$: in
first order,  $ \tilde\beta(t_1)=1$ and
  \BEq
\frac{{\tilde\a}(t_1)}{\tilde\b (t_1)} =\frac{-g_m}{\Delta_m(0)}
-ig_m \int_{0}^{t_1}e^{i{\cal B}_0^t}dt, \,\,\,
{\cal B}_{t'}^{t''} \equiv \int_{t'}^{t''}\D_{m}(t)\, dt .
 \label{alpha-beta-1}\EEq
Similarly, for an arbitrarily chosen rear ramp shape we know desired
$\tilde\alpha$ and $\tilde\beta$ at the end of the flat part $t_2$:
in first order,
$\tilde\a(t_2)= \tilde\a(t_{\rm f})= -ie^{-i\varphi}$ and
 \BEq
  \frac{{\tilde\b}(t_2)}{{\tilde\a}(t_2)} = \frac{g_m}{\Delta_m (t_{\rm f})} \,
e^{-i{\cal B}_0^{t_{\rm f}}} +
 ig_m e^{-i{\cal B}_0^{t_{\rm f}}}
\int_{t_2}^{t_{\rm f}}e^{i{\cal B}_{t}^{t_{\rm f}}} dt  .
  \label{alpha-beta-2}\EEq
During the flat part of the pulse we can use the two-level approximation with
coupling $g_m$, and essentially connect the two points on the Bloch sphere
corresponding to Eqs.\  (\ref{alpha-beta-1}) and (\ref{alpha-beta-2}) by a
``Rabi'' pulse.
 These points are close to the North and South poles, so the
pulse is close to the ideal $\pi$-pulse; we assume a small constant
overshoot $\D_m \equiv -D$ with $|D/g_m| \ll 1$ (Fig.\ 4), and
duration $t_2 - t_1 = \pi/\omega_R - \tau$ with $|\tau|\ll
\pi/\omega_R$, where $\w_R \equiv \sqrt{4g_m^2+D^2}$. Then using the
leading-order relation for an almost perfect $\pi$-pulse,
    \BEq
\label{eq:RabiApprox}
\frac{{\tilde\b}(t_2)}{{\tilde\a}(t_2)}=\frac{{\tilde\a}(t_1)}{{\tilde\b}(t_1)}
+\frac{D}{2g_m}+ig_m\tau, \,\,\,
\frac{{\tilde\a}(t_2)}{{\tilde\b}(t_1)}= -ie^{-i\pi D/4g_m} ,
    \EEq
we obtain the needed pulse parameters $D$ and $\tau$ as
\begin{align}
\label{eq:overshoot}
 \frac{D}{2g_m^2 } + i\tau &=
\frac{1}{\D_m(0)}
+ \frac{e^{-i{\cal B}_{0}^{t_{\rm f}}}}{\D_m(t_{\rm f})}
\nonumber \\
& + i \int_{0}^{t_1} e^{i{\cal B}_{0}^{t}} dt
+ i e^{-i{\cal B}_{0}^{t_{\rm f}}}
 \int_{t_{2}}^{t_{\rm f}} e^{i{\cal B}_{t}^{t_{\rm f}}} dt,
\end{align}
and also find the resulting phase $\varphi = \pi D/4g_m$.

    We have checked numerically the analytical pulse design given by
Eqs.\ (\ref{eq:shaping2}) and (\ref{eq:overshoot}).
    For example, for
a piecewise-linear pulse whose front ramp consists of two straight
segments (Fig.\ 4), the error
    \BEq
\label{eq:MOVEfidelity} {\rm Err} = 1- |\bra \overline{100}|U_{\rm
MOVE}|\overline{010}\ket|^2
    \EEq
for the analytically designed pulses is found to be below $10^{-3}$
for typical parameters with $g_m/2\pi=g_b/2\pi= 25$ MHz. As
expected, the numerical four-parameter optimization of such pulse
shape gives zero error (up to machine accuracy), and the shape of
this perfect pulse is close to the analytically-designed shape (see
Fig.\ 4).

%%%%%%%%%%%% FIGURE 5 %%%%%%%%%%%
\begin{figure}
\includegraphics[angle=0,width=1.00\linewidth]{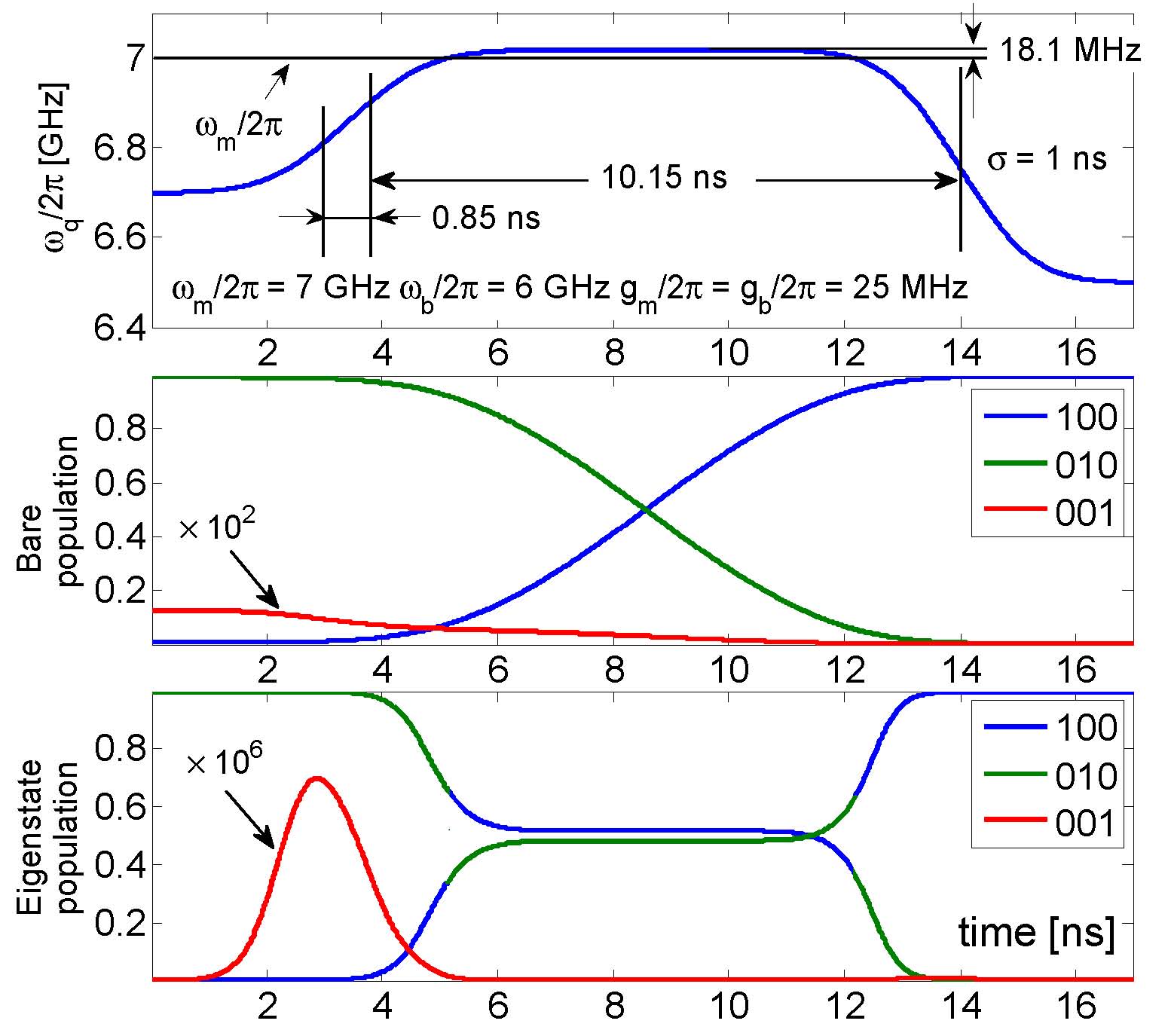}
\caption{ \label{fig:5} (Color online) Implementation of the
qubit$\rightarrow$memory MOVE operation in a three-component $mqb$
system using a pulse with error-function-shaped ramps (sum of two
time-shifted error functions for the front ramp). Four parameters of
the pulse shape (see upper panel) are optimized: the time shift and
amplitude ratio for the front-ramp error functions, the duration of
the middle part of the pulse, and the overshoot magnitude. The error
functions are produced by integrating Gaussians with standard
deviation $\sigma=1$ ns; the beginning and end of the pulse are at
3$\sigma$ from the nearest error-function centers (shown by vertical
lines in the upper panel). The middle and lower panels show
time-dependence of the level populations in the bare-state basis and
comoving eigenbasis. The MOVE error (\ref{eq:MOVEfidelity}) is zero
up to machine accuracy.}
\end{figure}
%%%%%%%%%%%% END FIG. 5 %%%%%%%%%%

    Experimental pulses for the MOVE operation
\cite{WANG2011,Mariantoni-NatPhys,Mariantoni-Science} are produced
by a Gaussian filter and therefore have the error-function-shape
ramps. We can use the same design idea for such pulses: shaping the
front ramp using two parameters (see Fig.\ \ref{fig:5}) takes care
of the``tail'' $\tilde\gamma(t_{\rm f})$ on the bus, for the rear
ramp we use any convenient shape, and for the middle part we vary
the overshoot frequency and duration to ensure proper population
transfer between the qubit and the memory (for such pulses it is
natural to define the duration to be between the inflection points
of the error-function shapes). We have checked numerically that
these four parameters are sufficient to achieve zero error (perfect
transfer fidelity $F=1-{\rm Err}$) in the truncated three-element
system.

    As a further simplification of the MOVE pulse design, let us optimize
only two middle-part parameters (overshoot and duration) and do not
optimize the front ramp shape. In this case we cannot ensure the
proper ``tail'' $\tilde\gamma(t_{\rm f})$; however, it is small by
itself, and therefore the error is not large. Moreover, for
sufficiently slow pulses the ``tail'' is almost correct
automatically because of the adiabatic theorem. It is important to
notice that the bus is well-detuned, $|\Delta_b/g_b|\gg 1$, and then
the adiabaticity condition is $| d\Delta_b/dt |\ll
|\Delta_b^3/g_b|$, which is well-satisfied even by rather fast
pulses. To estimate the corresponding error, we consider the
two-level system bus-memory during the front ramp and write the
differential equation for the variable $y\equiv \gamma/\beta
-g_b/\Delta_b$, which describes deviation from the co-moving
eigenstate. Assuming $|\gamma/\beta|\ll 1$ and $|\dot{\omega}_q
\gamma/\beta^2|\ll 1$, we obtain
   $\dot{y}=i\Delta_b y+g_b\Delta_b^{-2}\dot{\omega}_q$. The ``tail'' error at
the end of the front ramp is $|y(t_1)^2|$ (notice that $|\beta(t_1)|\approx 1$),
and it does not change significantly
during the rest of the pulse. If this is the major contribution to the MOVE
error, then
    \BEq
 {\rm Err} = \left| \int_0^{t_1} \frac{g_b}{\Delta_b^2(t)} \, \dot{\omega}_q(t)
 \exp(-i{\cal A}_0^t) \, dt \right|^2,
    \label{Err-tail}\EEq
where ${\cal A}_0^t$ is defined in Eq.\ (\ref{cal-A}).
Numerical optimization of only the middle part of the pulse (overshoot and
duration) confirms that Eq.\ (\ref{Err-tail}) is a good approximation for
the MOVE error in this case. Notice that for an error-function ramp obtained
by integrating a Gaussian with the standard deviation (time-width) $\sigma$,
the error (\ref{Err-tail}) decreases exponentially with $\sigma$ (we assume
sufficiently long ramp time $t_1$) and is typically quite small. For example,
for $\Delta_b/2\pi$ changing from 0.5 GHz to 1 GHz and $g_b/2\pi=50$ MHz, the
error is below $10^{-4}$ for $\sigma>0.5$ ns (for $\sigma>0.35$ ns if
$g_b/2\pi=25$ MHz).

    So far we have only considered the MOVE qubit$\rightarrow$memory. The
MOVE in the opposite direction memory$\rightarrow$qubit can be designed
by using the time-reversed pulse shape. Perfect MOVE still requires optimizing
four parameters (overshoot and duration of the middle part and the two
parameters
for the rear ramp), while using only the two parameters for the middle part
is sufficient for a high-fidelity MOVE.

 In designing the MOVEs between
the qubit and the memory we assumed no quantum information on the
bus. The presence of an excitation on the bus makes the previously
designed perfect MOVE imperfect. We checked numerically that the
corresponding error for typical parameters is about $10^{-4}$, i.e.\
quite small. Moreover, a typical RezQu algorithm never needs a MOVE
between a qubit and memory with occupied bus, so the unaccounted
MOVE errors due to truncation are even much smaller. The analysis in
this paper assumes the RWA Hamiltonian (\ref{RWAhamiltonian1}),
which neglects terms $\propto a_m^{\pm}\sigma_q^{\pm}$ and $\propto
a_b^{\pm}\sigma_q^{\pm}$, which change the number of excitations. We
have checked numerically that addition of these terms into the
Hamiltonian leads to negligibly small changes of the system dynamics
during the MOVE operations.

 Designing MOVEs between the qubit and the bus is similar to
designing MOVEs between the qubit and memory, if we consider the
truncated three-element system. However, in reality the situation is
more complicated because the bus is coupled with other qubits. Our
four-parameter argument in this case does not work, and designing a
perfect single-excitation MOVE would require $2N+2$ parameters (for
a truncated system with $N$ qubits, one memory, and the bus), which
is impractical. However, the occupation of additional qubits is
essentially the effect of the ``tails'' (if the discussed below
problem of level crossing is avoided). Therefore, the desired
``tails'' can be obtained automatically by using sufficiently
adiabatic ramps in the same way as discussed above (for a MOVE
qubit$\rightarrow$bus the front ramp will be important for the
``tails'' from both sides, i.e.\ on the memory and other qubits). In
analyzing the dynamics of the ``tails'' at other qubits, it is
useful to think it terms of eigenstates of a truncated system, which
includes the bus and other qubits (while excluding the qubit
involved in the MOVE). Then the ``tail'' error is the occupation of
the eigenstates, mainly localized on other qubits. Since the
frequencies of the bus and other qubits do not change with time, for
the error calculation it is still possible to use Eq.\
(\ref{Err-tail}), in which $\Delta_b$ is replaced with $\omega_q
-\omega_{q,k}$ (for the ``tail'' on $k$th qubit), $g_b$ is replaced
with $g_bg_{b,k}/(\omega_{q,k}-\omega_b)$, and ${\cal A}_0^t$ is
replaced with $\int_0^t [\omega_q (t')-\omega_{q,k}]\, dt'$, where
the subscript $k$ labels additional qubit. This gives the estimate
    \BEq
 {\rm Err} = \left| \int_0^{t_1} \frac{g_b g_{b,k}\, \dot{\omega}_q(t)}
 {\Delta_{q,k}^2(t) \, \Delta_{b,k}} \,
 e^{-i\int_0^{t} \Delta_{q,k}(t')\, dt'} \, dt \right|^2,
    \label{Err-tail-k}\EEq
of the error due to the ``tail'' on $k$th qubit for the
qubit$\rightarrow$bus MOVE, in which integration is within the front
ramp, $\Delta_{b,k}=\omega_{q,k}-\omega_b$, and
$\Delta_{q,k}(t)=\omega_q(t)-\omega_{q,k}$. The formula for the
bus$\rightarrow$qubit MOVE is similar, but the integration should be
within the rear ramp.
   The error (\ref{Err-tail-k}) should be summed
over additional $N-1$ qubits (index $k$) and therefore can be
significantly larger than in our calculations for a truncated $mqb$
system; however, the error increase is partially compensated by
smaller effective coupling $g_bg_{b,k}/\Delta_{b,k}$. Crudely, we
expect errors below $10^{-4}$ for smooth ramps of few-nanosecond
duration and $N<10^2$. We emphasize that this simple solution of the
``tail'' problem is possible only when we use eigenstates to
represent the logical states.

    Another problem which we did not encounter in the analysis of the
truncated system is the level crossing with other (empty) qubits
during the MOVE operation. A simple estimate of the corresponding
error is the following. Effective resonant coupling between the
moving qubit and another ($k$th qubit) via the bus is
$g_{b}g_{b,k}/\Delta_b$, where $g_{b}$ and $g_{b,k}$ are two
qubit-bus couplings and the detuning $\Delta_b$ is the same for both
qubits at the moment of level crossing. Then using the Landau-Zener
formula we can estimate the error (population of the other qubit
after crossing) as ${\rm Err} \simeq 2\pi
g_{b}^2g_{b,k}^2/(\Delta_b^2|\dot{\omega}_q|)$, where
$|\dot{\omega}_q|$ is the rate of the qubit frequency change at the
crossing. Our numerical calculations show that this estimate works
well, though up to a factor of about 2 [when curvature of
$\omega_q(t)$ at the point of crossing is significant]. Using this
estimate for $g_{b}/2\pi= g_{b,k}/2\pi= 25$ MHz, $\Delta_b/2\pi=500$
MHz, and $\dot{\omega}_q/2\pi = 500$ MHz/ns, we obtain a quite
significant error of about $10^{-4}$. A possible way to compensate
this error is by using interference of the Landau-Zener transitions
\cite{L-Z-compensation} on the qubit return transition. Another
solution of the problem is to park empty qubits outside the
frequency range between the bus and memories (above 7 GHz in our
example). This would make impossible to cancel the idling error of
Eq.\ (\ref{eq:OmegaZZ_4th_order}) by using the ``midway parking'',
but the idling error is still small even without this cancellation
[see the estimate below Eq.\ (\ref{eq:IE-RezQu})]. Besides the
qubit-qubit level crossings, there are also level crossings between
a moving qubit and other memories. This is a higher-order (weaker)
process because of three steps between the qubit and memory. The
effective coupling with $k$th memory is then $g_b
g_{b,k}g_{m,k}/(\Delta_b \Delta_{m,k})$, and the level crossing
error estimate is ${\rm Err} \simeq 2\pi
g_{b}^2g_{b,k}^2g_{m,k}^2/(\Delta_b^2\Delta_{m,k}^2
|\dot{\omega}_q|)$.

    In this paper we do not analyze two-qubit gates. Our preliminary
numerical simulation of the controlled-$Z$ gate has shown
possibility of a high-fidelity gate design (with the error of about
$10^{-3}$, mainly due to level crossing). However, we have not
studied this gate in detail. A detailed analysis of two-qubit gates
in the RezQu architecture will be presented elsewhere
\cite{JoydipGhosh11}.

\section{Tunneling measurement}
%\subsection*{Tunneling measurement}

Finally, let us discuss whether or not using the eigenstates as the
logical states presents a problem for measurement. Naively, one may think
about a projective measurement of an individual qubit; in this case the
logic state ``1'' would be erroneously measured as ``0'' with
probability of about $(g_m/\Delta_m)^2+(g_b/\Delta_b)^2\sim 10^{-2}$
because the
eigenstate spreads to the neighboring memory and bus. This would be a
very significant error, and the bare-state representation of logical
states would be advantageous. However, this is not actually the case
because any realistic measurement is not instantaneous (not projective).
In fact, if a measurement takes longer than $\Delta_{m,b}^{-1}$, then
the eigenbasis is better than the bare basis.

As a particular example let us analyze tunneling measurement of a
phase qubit \cite{Martinis-QIP} (we expect a similar result for the
qubit measurement in the circuit-quantum-electrodynamics (cQED)
setup \cite{Blais04,Girvin08}). The bare states $|0\rangle$ and
$|1\rangle$ of a phase qubit correspond to the two lowest energy
states in a quantum well, and the measurement is performed by
lowering the barrier separating the well from essentially a
continuum of states \cite{Martinis-QIP}. Then the state $|1\rangle$
tunnels into the continuum with a significant rate $\Gamma$, while
the tunneling rate for the state $|0\rangle$ is negligible. The
event of tunneling is registered by a detector ``click'' (the
detector is a SQUID, which senses the change of magnetic flux
produced by the tunneling). In the ideal case after waiting for a
time $t\gg \Gamma^{-1}$ the measurement error is negligibly small
(in real experiments the ratio of the two tunneling rates is only
$\sim 10^2$, which produces a few-per-cent error; however, we
neglect this error because in principle it can be decreased by
transferring the state $|1\rangle$ population to a higher level
before the tunneling, and also because here we are focusing on the
effect of ``tails'' in the neighboring elements).

    In presence of the memory and resonator coupled to the qubit,
the logic state ``0'' still cannot be misidentified, because the
tunneling is impossible without an excitation. However, the logic
state ``1'' can be misidentified as ``0'', when sometimes the
expected tunneling does not happen (because part of the excitation
is located in the memory and resonator). Let us find probability of
this error.
    For simplicity we consider a two-component model
in which a phase qubit is coupled to its memory resonator only, and
restrict the state space to the single-excitation subspace of this
$mq$ system. Then the tunneling process can be described by the
non-Hermitian Hamiltonian (e.g.,
\cite{GOTTFRIED_ParticlePhysicsIandII})
    \BEq
H =
\begin{bmatrix}
\w_m & g_m\cr
g_m & \w_q - i \Gamma /2
\end{bmatrix},
    \label{tun-H}\EEq
and the error in measuring the logic state ``1'' (identifying it as
``0'' after measurement for time $t$) is its survival probability
    \BEq \label{eq:tunnelingError} {\rm Err} =
|\bra\psi(t)|\psi(t)\ket|^2,
    \EEq
where the initial state is normalized,
$|\bra\psi(0)|\psi(0)\ket|^2=1$. Our goal is to compare this error
for the cases when the initial state is the bare state $|\psi(0)\ket
= |01\ket$ or the eigenstate $|\psi(0)\ket = |\overline{01}\ket$ (in
this $mq$ notation the qubit state is shown at the second place, and
$|\overline{01}\ket$ is the eigenstate before the measurement, i.e.\
when $\Gamma=0$).

    The solution $|\psi(t)\ket$ of the time-dependent
Schr\"{o}dinger equation,
$i(d/dt)|\psi(t)\ket  = H|\psi(t)\ket$, is given by the linear combination,
$|\psi(t)\ket = C_m |\widetilde{10}\ket \, e^{-iE_m t}
+ C_q |\widetilde{01}\ket \, e^{-iE_qt}$.
Here the eigenstate notation with the tilde sign reminds of a non-zero
$\Gamma$,  the constants $C_{m,q}$ depend on the initial conditions, and
$E_{m,q}\equiv {\rm Re}(E_{m,q}) -i\Gamma_{m,q}/2$
are the complex eigenenergies, which include the corresponding  decay
rates $\Gamma_{m}$ and $\Gamma_q$ of the eigenstates located mainly on
the memory and the qubit.
  Diagonalizing the Hamiltonian
(\ref{tun-H}) and assuming weak coupling, $g_m \ll \D_m$,
$g_m \ll \Gamma$, we find
    \BEq
\G_m = \frac{g_m^2\Gamma}{\D_m^2+(\Gamma/2)^2} \ll \Gamma, \quad
\G_q = \G - \G_m \approx \Gamma.
    \label{Gamma_mq}\EEq

%%%%%%%%%%% FIGURE 6 %%%%%%%%%%%
\begin{figure}
\includegraphics[angle=0,width=1.00\linewidth]{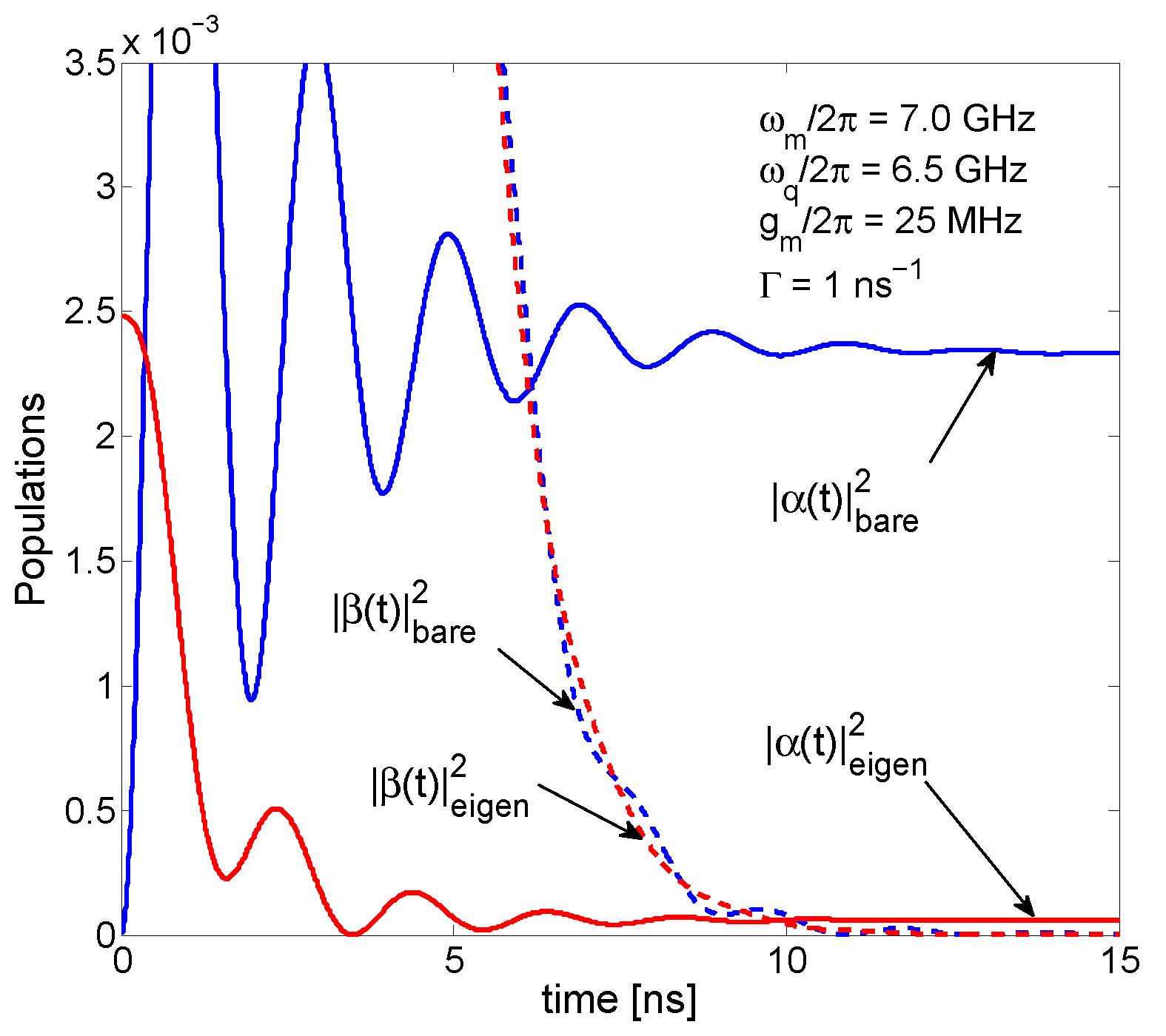}
\caption{ \label{fig:6} (Color online) Time dependence of the
squared amplitudes $|\a|^2$ and $|\b|^2$ of the $mq$ state
$|\psi(t)\rangle = \alpha (t) \, |10\rangle +\beta (t) \, |01\rangle
$, decaying in the process of tunneling measurement. Blue curves
correspond to the system initially prepared in the bare state
$|\psi(0)\rangle = |01\rangle$, while for red curves the initial
state is the eigenstate $|\psi(0)\rangle = |\overline{01}\rangle$.
For $t\gg \Gamma^{-1}$ the measurement error
(\ref{eq:tunnelingError}) is mainly the residual occupation
$|\a(t)|^2$ of the memory resonator (solid curves). For the depicted
system parameters $|\alpha_{\rm eigen}(t)|^2/|\alpha_{\rm
bare}(t)|^2 \approx 0.025$ at $t \agt 10$ ns, i.e.\ the error of the
eigenstate measurement is 40 times smaller than the error of the
bare state measurement. }
\end{figure}
%%%%%%%%%%% END FIG. 6 %%%%%%%%%%

For measurement during a sufficiently long time $t\gg \Gamma$,
only the $|\widetilde{10} \rangle$-term in $|\psi(t)\rangle$ survives,
and correspondingly the error (\ref{eq:tunnelingError}) is
${\rm Err}=|C_m|^2e^{-\Gamma_m t}$, where
$C_m=\langle \widetilde{10} |\psi(0)\rangle$. Thus we obtain
    \BEq
\frac{{\rm Err}_{\rm eigen}}{ {\rm Err}_{\rm bare}} =
\frac{\Gamma^2}{4\Delta_m^2}, \,\,\,
     {\rm Err}_{\rm bare}= \frac{e^{-\G_m t} g_m^2}{\D_m^2+(\Gamma/2)^2},
    \label{Err-meas}\EEq
for the measurement errors starting either with the eigenstate or with
the bare state. Even though both errors decrease with the measurement
time $t$ as $e^{-\Gamma_m t}$, the rate $\Gamma_m$ is small [see Eq.\
(\ref{Gamma_mq})], so for a realistically long measurement we can use
$e^{-\Gamma_m t}\approx 1$.

    Equation (\ref{Err-meas}) shows that from the measurement point of view
it is {\it advantageous to use the eigenstates} to represent the logical states
rather than the bare states  if $\Gamma <2\Delta_m$. For
a typical value $\Delta_m/2\pi=0.5$ GHz this requires $\Gamma^{-1}>0.16$ ns,
which is always the case.

    Figure \ref{fig:5} shows the state dynamics during the tunneling
measurement in the bare basis $|\psi(t)\rangle = \alpha (t) \, |10\rangle
+\beta (t) \, |01\rangle $, starting either with the eigenstate
$|\overline{01}\rangle$ or with the bare state $|01\rangle$. The oscillations
correspond to the beating frequency $\Delta_m/2\pi=0.5$ GHz. We see that
similarly to the above-analyzed dynamics in the eigenbasis,
$|\beta (t)|^2$ becomes exponentially small after $t\gg \Gamma^{-1}$, while
$|\alpha(t)|^2$ essentially saturates (decaying with a much smaller
rate $\Gamma_m$). Then for the assumed tunneling rate $\Gamma=1$ ns$^{-1}$
the ratio of errors ${\rm Err}_{\rm eigen}/{\rm Err}_{\rm bare}\approx
|\alpha_{\rm eigen}(t)|^2/|\alpha_{\rm bare}(t)|^2$  (with the subscript
denoting the initial state) saturates at approximately the value 0.025
given by Eq.\ (\ref{Err-meas}).

    We emphasize that even though we have only considered the tunneling
measurement, the result (\ref{Err-meas}) for the measurement error
is expected to remain crudely valid for most of realistic (i.e.\
``weak'') measurements with a time scale $\Gamma^{-1}$. In
particular, for the cQED setup we expect that the role of $\Gamma$
is played (up to a factor) by the ensemble dephasing rate due to
measurement.

\section{Conclusion}
%\subsection*{Summary}

%\vspace{0.5cm}

In summary, we have discussed the main ideas of the RezQu
architecture and analyzed several error mechanisms, excluding
analysis of two-qubit gates. The main advantage of the RezQu
architecture is the strong ($>10^4$ times) reduction of the idling
error compared to the conventional bus-based architecture, and also
an effective solution to most of the problems related to spectral
crowding. In the absence of decoherence this makes possible a simple
scaling of a RezQu device to $\sim 30$ qubits without the need for
dynamical decoupling. For further scaling the next architectural
level of communication between the RezQu devices seems to be needed.

    We have shown that instead of using bare states it is much better
to use eigenstates to represent logical states. In this case there
is essentially no dynamics in idling (except for the phase errors),
which greatly simplifies a modular construction of a quantum
algorithm. The logical encoding by eigenstates is also advantageous
for the single-qubit state generation and measurement. We have
presented a simple design for the MOVE operation, which is the most
frequent operation in the RezQu architecture. We have shown that a
four-parameter optimization is sufficient for designing a perfect
MOVE in a truncated three-component system. Moreover, optimization
of only two experimentally-obvious parameters is sufficient for
high-fidelity MOVEs (with errors less than $10^{-4}$). While we have
not analyzed two-qubit gates, we expect that their design with
similar high fidelity is also possible. Overall, we believe that the
RezQu architecture offers a very significant advantage compared to
the previously proposed architectures for superconducting qubits,
and we believe that this is the practical way to progress towards a
medium-scale quantum computing device.

% \acknowledgements
This work was supported by IARPA under ARO grant
W911NF-10-1-0334. The authors thank Michael Geller, Farid Khalili,
Matteo Mariantoni, Leonid Pryadko, and Frank Wilhelm for useful discussions.

\appendix
\section{DERIVATION OF $\Omega_{ZZ}$}
\label{sec:idlingErrorInRezQu}

In this Appendix we derive Eq.\ (\ref{eq:OmegaZZ_4th_order}) for
$\Omega_{ZZ}=\epsilon_{101}+\epsilon_{000}-\epsilon_{100}-\epsilon_{001}$
in the truncated $mqb$ system. We assume that the couplings $g_m$ and $g_b$
are of the same order, $g_m\sim g_b \sim g$, and do calculations in fourth
order in $g$.

  The RWA Hamiltonian (\ref{RWAhamiltonian1}) leads to
the formation of three subspaces, which do not interact with each
other: the ground state $|\overline{ 000}\rangle = |000\rangle$
(with zero energy, $\epsilon_{000}=0$), the single-excitation
subspace $\{|100\rangle, | 010\rangle, | 001\rangle\}$, and the
two-excitation subspace $\{|101\rangle, | 110\rangle, | 011\rangle,
|200\rangle, | 020\rangle, | 002\rangle\}$. It is rather easy to
find eigenenergies in the single-excitation subspace; neglecting the
direct coupling $g_d$, in fourth order in $g$ we obtain
\begin{align}
\label{eq:E100_4th}
\epsilon_{100} &= \w_m + \frac{g_m^2}{\Delta_m}
- \frac{g_m^4}{\Delta_m^2} + \frac{g_m^2g_b^2}{\Delta_m^2(\w_{m}-\w_b)},
    \\
\label{eq:E001_4th}
\epsilon_{001} &= \w_b - \frac{g_b^2}{\Delta_b}
- \frac{g_b^4}{\Delta_b^2} - \frac{g_m^2g_b^2}{\Delta_b^2(\w_{m}-\w_b)}.
\end{align}

To find $\e_{101}$, we write the eigenstate $|\overline{101}\rangle$
as a superposition of all elements of the two-excitation subspace,
    \begin{eqnarray}
&& |\overline{101}\ket = \left[ |{101}\ket + \a_{110} |110\ket + \a_{011} |011\ket
+ \a_{200} |200\ket \right. \quad
    \nonumber \\
&& \hspace{0.9cm}
\left. + \a_{020} |020\ket + \a_{002} |002\ket \right] /{\rm Norm}
    \end{eqnarray}
with unimportant normalization. Then the Schr\"{o}dinger equation
$H |\overline{101}\rangle=\epsilon_{101}|\overline{101}\rangle $
(again neglecting $g_d$) gives six equations:
\begin{align}
\label{eq:101system}
&(\w_m + \w_b) + \a_{110} g_b + \a_{011}g_m = \e_{101}, \nonumber \\
&g_b + \a_{110}(\w_m + \w_{q}) + (\a_{200} + \a_{020})g_m\sqrt{2}
= \a_{110} \e_{101}, \nonumber \\
&g_m + \a_{011}(\w_{q} + \w_b) + (\a_{020} + \a_{002})g_b\sqrt{2}
= \a_{011} \e_{101}, \nonumber \\
&\a_{110}g_m \sqrt{2} + \a_{200} 2 \w_m = \a_{200} \e_{101}, \nonumber \\
&\a_{110}g_m \sqrt{2} + \a_{011} g_b \sqrt{2} + \a_{020}(2\w_q - \eta)
= \a_{020} \e_{101}, \nonumber \\
&\a_{011}g_b \sqrt{2} + \a_{002} 2 \w_b = \a_{002} \e_{101}.
\end{align}
From the first three of them we obtain
\begin{align}
\label{eq:E101_intermediate}
\e_{101} =&
\w_m + \w_b
 + \frac{g_m^2 + (\a_{020}+\a_{002})g_mg_b\sqrt{2} }{\e_{101} - \w_q - \w_b}
\nonumber \\
& + \frac{g_b^2 + (a_{200}+\a_{020})g_mg_b\sqrt{2} }{\e_{101} - \w_m - \w_q},
\end{align}
which gives $\epsilon_{101}$ in fourth order in $g$ if we
use
second-order $\epsilon_{101}$ in the denominators (which is obtained
from the same equation using zeroth-order $\epsilon_{101}$) and
second-order amplitudes $\alpha_{200}$, $\alpha_{020}$, $\alpha_{002}$.
These amplitudes can be found from the last three equations (\ref{eq:101system})
using the first-order values $\a_{110} = -g_b/\Delta_b$,
$\a_{011}=g_m/\Delta_m$:
\begin{align}
& \a_{200}= \frac{g_mg_b\sqrt{2}}{\Delta_b(\w_m-\w_b)}, \,\,\,
\a_{002}= \frac{g_mg_b\sqrt{2}}{\Delta_m(\w_m-\w_b)}
\nonumber \\
& \a_{020}= \frac{-g_mg_b\sqrt{2}}{\Delta_m\Delta_b}\,
\frac{\w_m+\w_b-2\w_q}{\w_m+\w_b-(2\w_q - \eta)}.
\label{eq:2-state_amplitudes_2nd_order}\end{align}
Finally, substituting Eq.\ (\ref{eq:2-state_amplitudes_2nd_order}) into
Eq.\ (\ref{eq:E101_intermediate}), and using
Eqs.\ (\ref{eq:E100_4th}), (\ref{eq:E001_4th}), we obtain
Eq.\ (\ref{eq:OmegaZZ_4th_order}) for $\Omega_{ZZ}$ in fourth order.

    The above was the formal derivation of  Eq.\ (\ref{eq:OmegaZZ_4th_order}).
Let us also obtain it approximately. Since in a linear system
$\Omega_{ZZ}=0$ (excitations do not interact with each other), a
non-zero value can come only from the qubit nonlinearity $\eta$.
Assuming small $\eta$, we can use the first-order perturbation
theory in $\eta$ to find the energy shift of
$|\overline{101}\rangle$ due to contribution from $|020\rangle$
(occupation of the qubit second level),
    \BEq
 \Omega_{ZZ}=\delta\epsilon_{101}=
-\eta\, |\alpha_{020}|^2.
    \label{omega_zz-eta}\EEq
To find $\alpha_{020}$ we start with the first-order (in $g$)
eigenstate $|\overline{101}\rangle=|101\rangle
-(g_b/\Delta_b)\,|110\rangle+(g_m/\Delta_m)\,|011\rangle$ and then
obtain the next-order estimate
$\a_{020}=\sqrt{2}g_mg_b[1/\Delta_b-1/\Delta_m]/
(2\omega_q-\eta-\omega_m-\omega_b)$ [which coincides with Eq.\
(\ref{eq:2-state_amplitudes_2nd_order})]. If this estimate is
substituted into Eq.\ (\ref{omega_zz-eta}), then for $\Omega_{ZZ}$
we obtain Eq.\ (\ref{eq:OmegaZZ_4th_order}) with the squared second
fraction. However, if in the above formula for $\alpha_{020}$ we
neglect $\eta$ (as we should for first-order perturbation in
$\eta$), then we obtain Eq.\ (\ref{eq:OmegaZZ_4th_order}) with the
second fraction replaced by 1. One may say that the average between
these two results for small $\eta$ reproduces Eq.\
(\ref{eq:OmegaZZ_4th_order}); however, it is more appropriate to say
that this approximation (first order in $\eta$) can accurately
reproduce only the first fraction in Eq.\
(\ref{eq:OmegaZZ_4th_order}), while the second fraction is beyond
the accuracy of the approximation.

    A slightly different derivation reproduces Eq.\
(\ref{eq:OmegaZZ_4th_order}) exactly. Instead of using the
first-order approximation in $\eta$ [Eq.\ (\ref{omega_zz-eta})], let
us find the change of $\epsilon_{101}$ due to its repulsion from
$\epsilon_{020}$. Since the effective interaction is $g_{\rm
eff}=\sqrt{2}g_m g_b[1/\Delta_b-1/\Delta_m]$ (see the above estimate
of $\alpha_{020}$), the repulsion is $\delta\epsilon_{101}=-g_{\rm
eff}^2/ (2\omega_q-\eta-\omega_m-\omega_b)$. The difference between
this repulsion with and without nonlinearity $\eta$ gives
$\Omega_{ZZ}=-g_{\rm eff}^2
[(2\omega_q-\eta-\omega_m-\omega_b)^{-1}-
(2\omega_q-\omega_m-\omega_b)^{-1}]$, which reproduces Eq.\
(\ref{eq:OmegaZZ_4th_order}).

    These simple derivations do not change if we take into account
the direct interaction $g_d$ in the Hamiltonian (\ref{RWAhamiltonian1}),
which is of the second order, $g_d\sim g^2$, because of Eq.\ (\ref{g_d}).
Therefore, the fourth-order result (\ref{eq:OmegaZZ_4th_order}) for
$\Omega_{ZZ}$ should not change either. The rigorous fourth-order
calculation shows that $\epsilon_{100}+\epsilon_{001}$ increases
by $2g_dg_mg_b/\Delta_m\Delta_b$ due to $g_d$, but $\epsilon_{101}$
increases by the same amount, so that these contributions to $\Omega_{ZZ}$
cancel each other.


\begin{thebibliography}{}

\bibitem{CLARKE2008}
J. Clarke and F. K. Wilhelm, Nature {\bf 453}, 1031 (2008).

\bibitem{Yamamoto03}
T. Yamamoto, Yu. A. Pashkin, O. Astafiev, Y. Nakamura, \& J. S.
Tsai, Nature {\bf 425}, 941 (2003).

\bibitem{Plantenberg07}
J. H. Plantenberg, P. C. de Groot, C. J. P. M. Harmans \& J. E.
Mooij, Nature {\bf 447}, 836 (2007).

\bibitem{Steffen06}
M. Steffen, M. Ansmann, R. C. Bialczak, N. Katz, E. Lucero, R.
McDermott, M. Neeley, E. M. Weig, A. N. Cleland and J. M. Martinis,
Science {\bf 313}, 1423 (2006).

\bibitem{DiCarlo09}
L. DiCarlo, J. M. Chow, J. M. Gambetta, L. S. Bishop, B. R. Johnson,
D. I. Schuster, J. Majer, A. Blais, L. Frunzio, S. M. Girvin \& R.
J. Schoelkopf, Nature {\bf 460}, 240 (2009).

\bibitem{DiCarlo10}
L. DiCarlo, M. D. Reed, L. Sun, B. R. Johnson, J. M. Chow, J. M.
Gambetta, L. Frunzio, S. M. Girvin, M. H. Devoret \& R. J.
Schoelkopf, Nature {\bf 467}, 574 (2010).

\bibitem{Mariantoni-Science}
M. Mariantoni, H. Wang, T. Yamamoto, M. Neeley, R. C. Bialczak, Y.
Chen, M. Lenander, Erik Lucero, A. D. O'Connell, D. Sank, M. Weides,
J. Wenner, Y. Yin, J. Zhao, A. N. Korotkov, A. N. Cleland, J. M.
Martinis, Science {\bf 334}, 61 (2011).

\bibitem{Clarke-06} T. Hime, P. A. Reichardt, B. L. T. Plourde,
T. L. Robertson, C. E. Wu, A. V. Ustinov, and J. Clarke, Science
{\bf 314}, 1427 (2006).

\bibitem{Simmonds-10}
 F. Altomare, J. I Park, K. Cicak, M. A. Sillanpaa, M. S. Allman, D. Li,
 A. Sirois, J. A. Strong, J. D. Whittaker, and R. W. Simmonds,
 Nature Phys. {\bf 6},  777 (2010).

\bibitem{Oliver-11}
J. Bylander, S. Gustavsson, F. Yan, F. Yoshihara, K. Harrabi, G.
Fitch, D. G. Cory, Y. Nakamura, J. S. Tsai, and W. D. Oliver, Nature
Phys. {\bf 7}, 565 (2011).

\bibitem{Esteve-09} F. Mallet, F. R. Ong, A. Palacios-Laloy, F. Nguyen,
P. Bertet, D. Vion, and D. Esteve, Nature Phys. {\bf 5}, 791 (2009).

\bibitem{Siddiqi-11}
R. Vijay, D. H. Slichter, and I. Siddiqi, Phys. Rev. Lett. {\bf
106}, 110502 (2011).


\bibitem{DiVincenzo} For a high-level architecture solution,
see D. P. DiVincenzo, Phys. Scr. {\bf T137}, 014020 (2009).

\bibitem{Martinis-QIP} J. M. Martinis, Quant. Inf. Processing {\bf 8},
81 (2009).

\bibitem{WANG2011}
H. Wang, M. Mariantoni, R. C. Bialczak, M. Lenander, E. Lucero, M. Neeley, A. O'Connell, D. Sank, M. Weides, J. Wenner, T. Yamamoto, Y. Yin, J. Zhao, J. M. Martinis, A. N. Cleland, Phys. Rev. Lett. {\bf 106}, 060401 (2011).

\bibitem{Mariantoni-NatPhys}
M. Mariantoni, H. Wang, R. C. Bialczak, M. Lenander, E. Lucero, M. Neeley, A. D. O'Connell, D. Sank, M. Weides, J. Wenner, T. Yamamoto, Y. Yin, J. Zhao, J. M. Martinis, A. N. Cleland, Nature Phys. {\bf 7}, 287 (2011).


\bibitem{WILHELM_NOON_2010}
S. T. Merkel and F. K. Wilhelm, New J. Phys. {\bf 12}, 093036 (2010).

\bibitem{Strauch03}
F. W. Strauch, P. R. Johnson, A. J. Dragt, C. J. Lobb, J. R. Anderson, and F. C. Wellstood, Phys. Rev. Lett. {\bf 91}, 167005 (2003).

\bibitem{Haack10}
G. Haack, F. Helmer, M. Mariantoni, F. Marquardt, and E. Solano, Phys. Rev. B {\bf 82}, 024514 (2010).

\bibitem{Yamamoto10}
T. Yamamoto, M. Neeley, E. Lucero, R. C. Bialczak, J. Kelly, M. Lenander, M. Mariantoni, A. D. O’Connell, D. Sank, H. Wang, M. Weides, J. Wenner, Y. Yin, A. N. Cleland, and J. M. Martinis, Phys. Rev. B {\bf 82}, 184515 (2010).

\bibitem{Nakamura-07-tunable}
A. O. Niskanen, K. Harrabi, F. Yoshihara, Y. Nakamura, S. Lloyd, and
J. S. Tsai, Science {\bf 316}, 723 (2007).

\bibitem{BIALCZAK2010}
R. C. Bialczak, M. Ansmann, M. Hofheinz, E. Lucero, M. Neeley, A. D. O'Connell, D. Sank, H. Wang, J. Wenner, M. Steffen, A. N. Cleland, J. M. Martinis, Nature Physics {\bf 6}, 409 (2010).

%
%\bibitem{JCPaper}
%E. T. Jaynes, F. W. Cummings, Proc. IEEE {\bf 51}(1), 89  (1963).

%%%%%%

\bibitem{Zhu03}
S.-L. Zhu, Z. D. Wang, and K. Yang, Phys. Rev. A {\bf 68}, 034303 (2003).

\bibitem{Blais03}
A. Blais, A. M. van den Brink, and A. M. Zagoskin, Phys. Rev.
Lett. {\bf 90}, 127901 (2003).

\bibitem{Zhou04}
X. Zhou, M. Wulf, Z. Zhou, G. Guo, and M. J. Feldman,
Phys. Rev. A {\bf 69}, 030301 (2004).


%%%%%%

\bibitem{Blais04}
A. Blais, R.-S. Huang, A. Wallraff, S. M. Girvin, and R. J. Schoelkopf, Phys. Rev. A {\bf 69}, 062320 (2004).

\bibitem{Cleland04}
A. N. Cleland and M. R. Geller, Phys. Rev. Lett. {\bf 93}, 070501 (2004).


\bibitem{Blais07}
A. Blais, J. Gambetta, A. Wallraff, D. I. Schuster, S. M. Girvin, M. H. Devoret, and R. J. Schoelkopf, Phys. Rev. A {\bf 75}, 032329 (2007).

\bibitem{Koch07}
J. Koch, T. M. Yu, J. Gambetta, A. A. Houck, D. I. Schuster, J. Majer, A Blais, M. H. Devoret,
S. M. Girvin, and R. J. Schoelkopf, Phys. Rev. A {\bf 76}, 042319 (2007).

\bibitem{MAJER2007}
J. Majer, J. M. Chow, J. M. Gambetta, Jens Koch, B. R. Johnson, J. A. Schreier, L. Frunzio, D. I. Schuster, A. A. Houck, A. Wallraff, A. Blais,
M. H. Devoret, S. M. Girvin \& R. J. Schoelkopf,
Nature {\bf 449}, 443 (2007).

\bibitem{Girvin08}
R. J. Schoelkopf, S. M. Girvin, Nature {\bf 451}, 664 (2008).


\bibitem{NORI11}
J. Q. You, F. Nori, Nature, {\bf 474}, 589 (2011).

%%%%%


\bibitem{Pritchett05}
E. J. Pritchett and M. R. Geller, Phys. Rev. A {\bf 72}, 010301(R) (2005).

\bibitem{Silanpaa07}
M. A. Sillanp\"{a}\"{a}, J. I. Park, and R. W. Simmonds,
Nature {\bf 449}, 438 (2007).

\bibitem{Johnson10}
B. R. Johnson, M. D. Reed, A. A. Houck, D. I. Schuster, L. S.
Bishop, E. Ginossar, J. M. Gambetta, L. DiCarlo, L. Frunzio, S. M.
Girvin \& R. J. Schoelkopf, Nature Physics {\bf 6}, 663 (2010).

\bibitem{JoydipGhosh11}
J.\ Ghosh et al., {\it Controlled-Z logic gate in
Resonator/zero-Qubit architecture}, in preparation (2011).


\bibitem{NIELSEN&CHUANG}
M. A. Nielsen and I. L. Chuang, {\it Quantum Computation and
Quantum Information} (Cambridge University Press, Cambridge,
UK, 2000).


\bibitem{Boissonneault09}
M. Boissonneault, J. M. Gambetta, and A. Blais,
Phys. Rev. A {\bf 79}, 013819 (2009).


\bibitem{Pinto10} R. A. Pinto, A. N. Korotkov, M. R. Geller, V. S. Shumeiko,
and J. M. Martinis, Phys. Rev. B {\bf 82}, 104522 (2010).

\bibitem{Hofheinz}
M. Hofheinz, H. Wang, M. Ansmann, R. C. Bialczak, Erik Lucero, M.
Neeley, A. D. O'Connell, D. Sank, J. Wenner, J. M. Martinis \& A. N.
Cleland, Nature {\bf 459}, 546 (2009).


\bibitem{PRYADKO&KOROTKOV_iSWAP}
L. P. Pryadko and A. N. Korotkov, unpublished.



\bibitem{L-Z-compensation}
W. D. Oliver, Yang Yu, J. C. Lee, K. K. Berggren, L. S. Levitov and
T. P. Orlando, Science {\bf 310}, 1653 (2005); M. Sillanp\"a\"a, T.
Lehtinen, A. Paila, Yu. Makhlin, and P. Hakonen, Phys. Rev. Lett.
{\bf 96}, 187002 (2006).

\bibitem{GOTTFRIED_ParticlePhysicsIandII}
K. Gottfried, V. F. Weisskopf, {\it Concepts of Particle Physics}, Vol.\ I
(Oxford University Press, 1984), p.\ 151.

\end{thebibliography}
\end{document}